\documentclass[prl, twocolumn, showpacs]{revtex4}
\usepackage{graphics,graphicx}
\begin{document}
\title{Dynamic Scaling in the Susceptibility of the Spin-$\frac{1}{2}$ Kagome Lattice \\ Antiferromagnet Herbertsmithite}
\author{J.S.~Helton$^{1,3}$}
\author{K.~Matan$^{1a}$}
\author{M.P.~Shores$^{2b}$}
\author{E.A.~Nytko$^{2}$}
\author{B.M.~Bartlett$^{2c}$}
\author{Y.~Qiu$^{3,4}$}
\author{D.G.~Nocera$^{2}$}
\author{Y.S.~Lee$^{1*}$}
\affiliation{$^{1}$Department of Physics, Massachusetts Institute
of Technology, Cambridge, MA 02139} \affiliation{$^{2}$Department
of Chemistry, Massachusetts Institute of Technology, Cambridge, MA
02139} \affiliation{$^{3}$NIST Center for Neutron Research, National Institute of Standards and Technology,
Gaithersburg, MD 20899} \affiliation{$^{4}$Department of Materials
Science and Engineering, University of Maryland, College Park, MD
20742}
\date{\today}

\begin{abstract}

The spin-$\frac{1}{2}$ kagome lattice antiferromagnet
herbertsmithite, ZnCu$_{3}$(OH)$_{6}$Cl$_{2}$, is a candidate
material for a quantum spin liquid ground state. We show that the
magnetic response of this material displays an unusual scaling
relation in both the bulk ac susceptibility and the low energy
dynamic susceptibility as measured by inelastic neutron scattering.
The quantity $\chi T^\alpha$ with $\alpha \simeq 0.66$ can be
expressed as a universal function of $H/T$ or $\omega/T$. This
scaling is discussed in relation to similar behavior seen in
systems influenced by disorder or by the proximity to a quantum
critical point.

\end{abstract}

\pacs{75.40.Gb, 75.50.Ee, 78.70.Nx} \maketitle

A continuing challenge in the field of frustrated magnetism is the
search for candidate materials which display quantum disordered
ground states in two dimensions. In recent years, a great deal of
attention has been given to the spin-$\frac{1}{2}$ nearest-neighbor
Heisenberg antiferromagnet on the kagome lattice, consisting of
corner sharing triangles.  Given the high frustration of the
lattice and the strength of quantum fluctuations arising from
spin-$\frac{1}{2}$ moments, this system is a very promising
candidate to display novel magnetic ground states, including the
``resonating valence bond" (RVB) state proposed by
Anderson\cite{Anderson1987}.  A theoretical and
numerical consensus has developed that the ground state of this system is not
magnetically
ordered\cite{ZengElser,MarstonZeng,SinghHuse,Sachdev1992,Lecheminant,Waldtmann,Mila},
although the exact ground state is still a matter of some debate.
Experimental studies of this system have long been hampered by a
lack of suitable materials displaying this motif.

The mineral herbertsmithite\cite{Shores,Braithwaite},
ZnCu$_{3}$(OH)$_{6}$Cl$_{2}$, is believed to be an excellent realization
of a spin-$\frac{1}{2}$ kagome lattice antiferromagnet.  The
material consists of kagome lattice planes of spin-$\frac{1}{2}$
Cu$^{2+}$ ions.  The superexchange interaction between nearest-neighbor spins leads to an antiferromagnetic coupling of $J = 17
\pm 1$~meV.  Extensive measurements on powder samples of
herbertsmithite have found no evidence of long range magnetic order
or spin freezing to temperatures of roughly 50~mK\cite{Helton,Mendels,Ofer}. Magnetic excitations are effectively
gapless, with a Curie-like susceptibility at low temperatures.  The
magnetic kagome planes are separated by layers of nonmagnetic
Zn$^{2+}$ ions; however, it has been suggested that there could be
some site disorder between the Cu and Zn ions\cite{Misguich,deVries1}.
This possible site disorder, with $\approx$~5\% of the magnetic Cu ions
residing on out-of-plane sites, as well as the presence of a
Dzyaloshinskii-Moriya (DM) interaction\cite{RigolPRB}, would likely influence the
low energy magnetic response.

In this Letter we report a dynamic scaling analysis of the
susceptibility of herbertsmithite as measured in both the bulk ac
susceptibility and the low energy dynamic susceptibility measured
by inelastic neutron scattering.  In particular, we find that the
quantity $\chi T^{\alpha}$ can be expressed as a universal function
in which the energy or field scale is set only by the temperature.
This type of scaling behavior, when measured in quantum
antiferromagnets\cite{SachdevYe} and heavy-fermion
metals\cite{ColemanPepin}, has long been associated with proximity
to a quantum critical point (QCP).  Power law signatures in the
susceptibility have also been
associated with random systems such as Griffiths
phase\cite{CastroNeto1998} or random singlet
phase\cite{BhattLee} systems.  Such similarities could shed light on
the relevant low energy interactions in herbertsmithite.

\begin{figure}
\centering
\includegraphics[width=8.0cm]{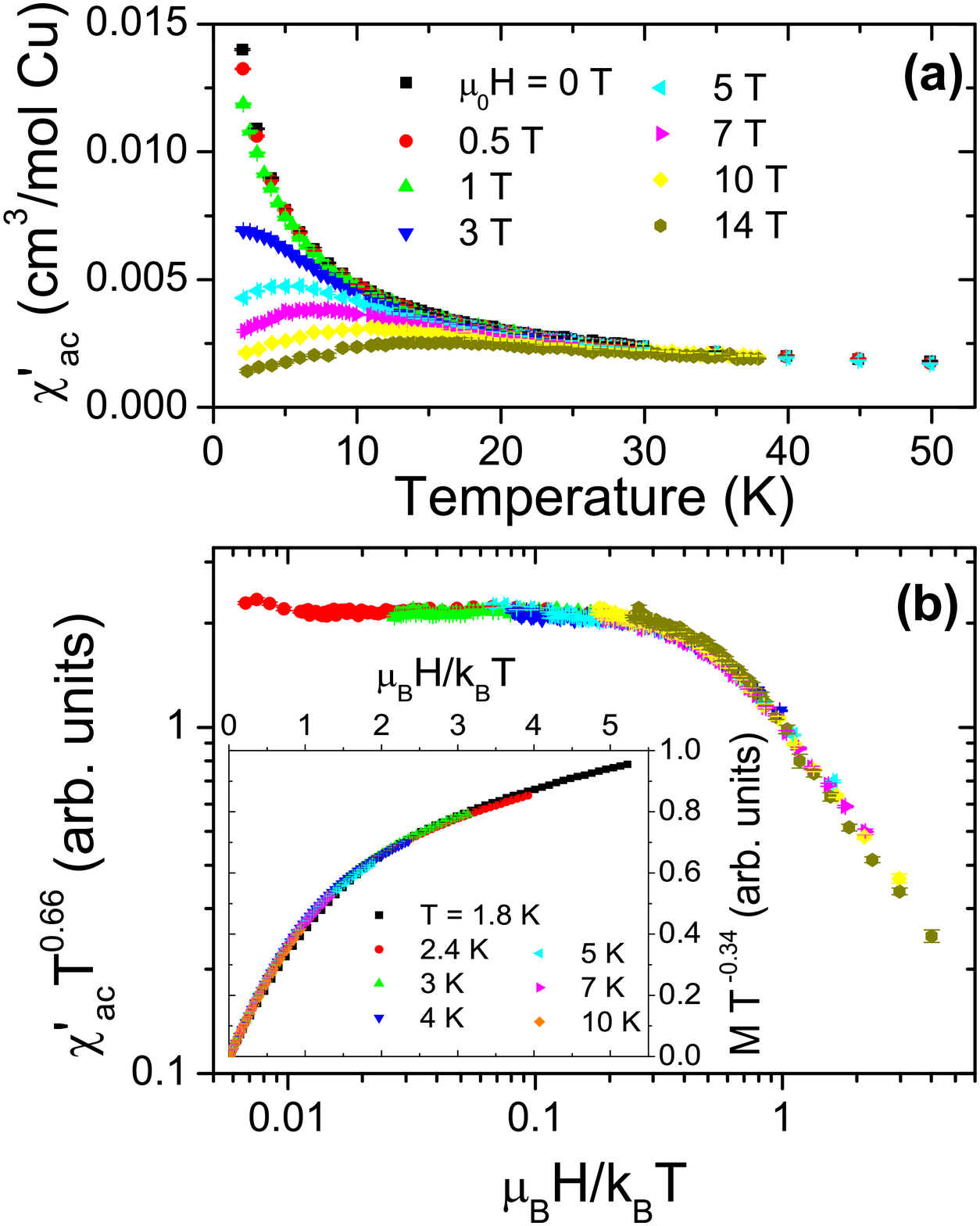}\vspace{-6mm}
\caption{(color online) (a) The in-phase component of the ac
susceptibility, measured at 100~Hz with an oscillating field of 17~Oe.  (b) A scaled plot of the ac susceptibility data measured at nonzero applied field,
plotted as $\chi^{\prime}_{ac}T^{\alpha}$ with $\alpha = 0.66$ on
the $y$ axis and $\mu_{B}H/k_{B}T$ on the $x$ axis. Inset: A scaled plot of the dc magnetization, showing $MT^{-0.34}$ vs $\mu_{B}H/k_{B}T$.
}\vspace{-6mm}
\label{ChiAC}
\end{figure}

Figure~\ref{ChiAC}(a) shows the ac magnetic susceptibility of a
herbertsmithite powder sample as measured using a commercial ac
magnetometer (Quantum Design).  An oscillating field of 17~Oe, with a frequency of
100 Hz, was applied along with a range of dc fields up to $\mu_{0}H = 14$~T. The data were corrected for the diamagnetic contribution by use of Pascal's constants.  These results, for data sets with nonzero applied dc field, are
plotted in Fig.~\ref{ChiAC}(b) with $\chi^{\prime}T^{\alpha}$ (with
$\alpha = 0.66$) on the $y$ axis and the unitless ratio
$\mu_{B}H/k_{B}T$ on the $x$ axis.  For this value of $\alpha$, the
data collapse quite well onto a single curve for a range of
$\mu_{B}H/k_{B}T$ spanning well over two decades.  Scaling plots with various exponent choices support $\alpha = 0.66 \pm 0.02$.  This scaling
remains roughly valid up to moderate temperatures, dependent upon
the applied dc field. In the data taken with $\mu_{0}H=0.5$~T, shown in
Fig.~\ref{ChiAC}, the scaling fails for temperatures greater than
roughly $T=35$~K; under an applied field of $\mu_{0}H=5$~T the scaling
remains valid to about $T=55$~K. The functional form of this
collapse is qualitatively similar to the generalized critical
Curie-Weiss function seen in the heavy-fermion compound
CeCu$_{5.9}$Au$_{0.1}$\cite{SchroderNature}, but with deviations
demonstrating that such a simple response function is not quite
adequate.  It should also be pointed out that in herbertsmithite
the entire bulk susceptibility obeys this scaling relation, while in
CeCu$_{5.9}$Au$_{0.1}$ it is only the estimated local contribution,
$\chi_{L}(T)$ = [$\chi(T)^{-1} - \chi(T=0)^{-1}$]$^{-1}$, that
obeys scaling.  A susceptibility of this form will imply a similar scaling in the bulk dc magnetization of the sample, with $MT^{\alpha-1}$ expressible as a function of $H/T$.  As a complementary measurement, such a scaling is shown in the inset to Fig.~\ref{ChiAC}(b).  The dc magnetization was measured up to $\mu_{0}H = 14$~T at temperatures ranging from $T = 1.8$~K to 10~K, and is plotted as $MT^{-0.34}$ vs $\mu_{B}H/k_{B}T$.

\begin{figure}
\centering
\includegraphics[width=7.4cm]{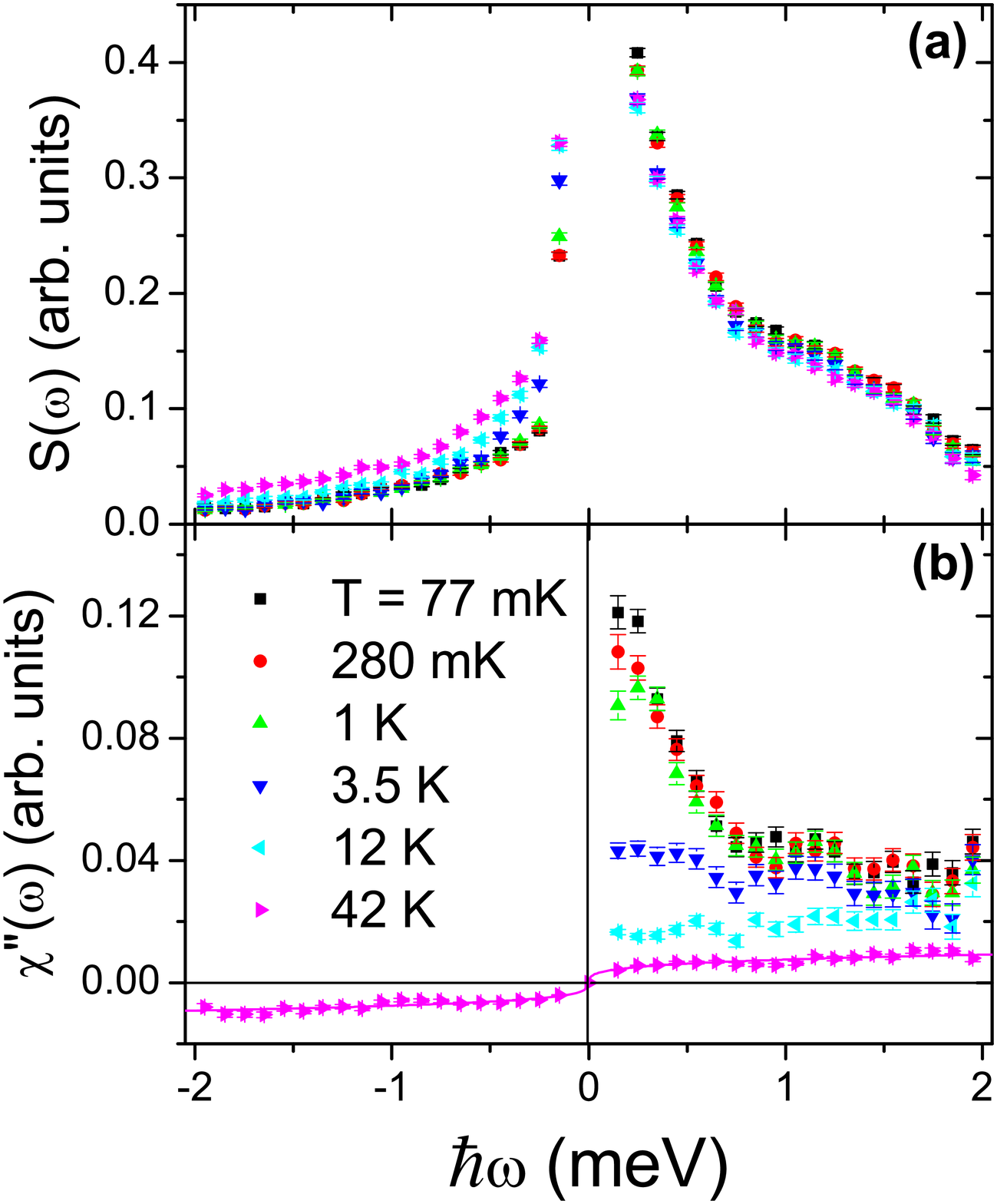}\vspace{-6mm}
\caption{(color online) (a) Neutron scattering structure factor $S(\omega)$, measured using DCS integrated over wavevectors $ 0.5 \leq Q \leq 1.9$~{\AA}$^{-1}$.  (b) The local dynamic susceptibility
$\chi''(\omega)$, determined as described in the text. Uncertainties where indicated in this article are statistical in origin and represent 1 standard deviation.}\vspace{-6mm}
\label{ChiPP}
\end{figure}

The inelastic neutron scattering spectrum of herbertsmithite was
measured on the time-of-flight Disk Chopper Spectrometer (DCS) at
the NIST Center for Neutron Research.  A deuterated powder sample of
mass 7.5~g was measured using a dilution refrigerator with an
incident neutron wavelength of 5~{\AA}. Measurements were taken at
six different temperatures, with roughly logarithmic spacing,
ranging from 77~mK to 42~K.  The scattering data were integrated over
a wide range of momentum transfers, $0.5 \leq  Q  \leq 1.9$~{\AA}$^{-1}$, to give a measure of the local response.  The momentum
integrated dynamic scattering structure factor, $S(\omega)$, is
shown in Fig.~\ref{ChiPP}(a).  Similar to previous reports on the
neutron scattering spectrum of herbertsmithite\cite{Helton}, the
data show a broad inelastic spectrum with no discernable spin
gap and only a weak temperature dependence for positive energy
transfer scattering. The negative energy transfer scattering
intensity is suppressed at low temperatures due to detailed balance.
The imaginary part of the dynamic susceptibility is related to the
scattering structure factor through the fluctuation-dissipation
theorem, $ \chi''(\omega) \, = \,
S(\omega)(1-e^{-\hbar\omega/k_{B}T})$.  The dynamic susceptibility can then be determined in a
manner similar to that used previously\cite{Helton}.  For the two lowest
temperatures measured, detailed balance considerations will
effectively suppress scattering at negative energy transfer for
values of $|\hbar\omega| \geq 0.15$~meV.  Thus these data sets
are averaged together and treated as background.  This background is
subtracted from the $T=42$~K data, for which the detailed balance
suppression is not pronounced below $|\hbar\omega| = 2$~meV. From
this, $\chi''(\omega$; $T=42$~K) is calculated for negative
$\omega$, and the values for positive $\omega$ are easily determined
from the fact that $\chi''(\omega)$ is an odd function of $\omega$.
The dynamic susceptibility at the other temperatures is calculated by
determining the difference in scattering intensity relative to the
$T=42$~K data set.  It is reasonably assumed that the elastic incoherent
scattering and any other background scattering are effectively
temperature independent. The calculated values
of $\chi''(\omega)$ at all measured temperatures are shown in Fig.~\ref{ChiPP}(b).  The $T=42$~K scattering data and
$\chi''(\omega)$ were fit to smooth functions for use in calculating
the susceptibility at other temperatures so that statistical errors
would not be propagated throughout the data; the smooth function of $\chi''(\omega$; $T=42$~K) used in the calculation is also shown in the figure.

\begin{figure}
\centering
\includegraphics[width=9.0cm]{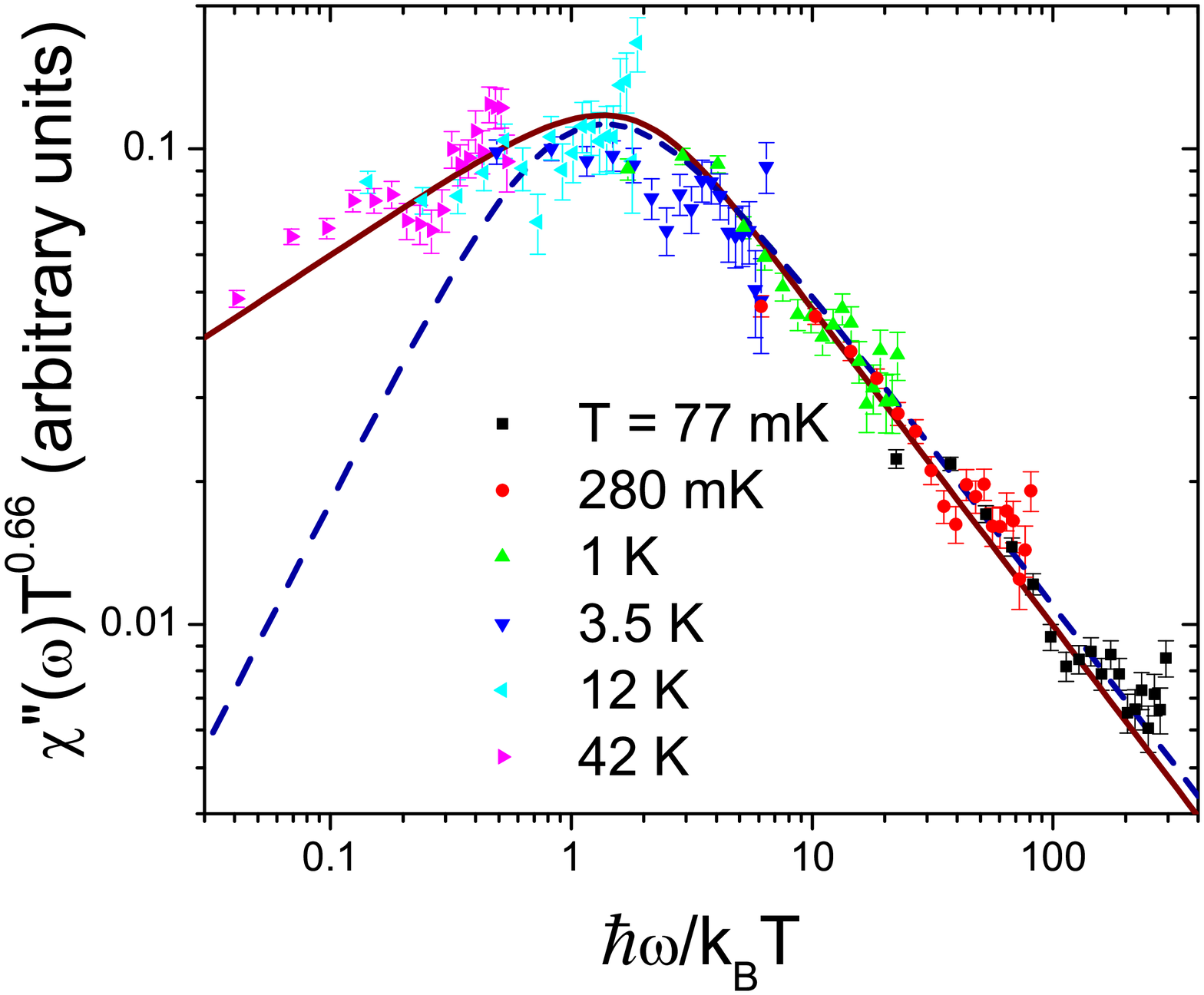}\vspace{-4mm}
\caption{(color online) The quantity $\chi''(\omega)T^{\alpha}$ with $\alpha =
0.66$ plotted against $\hbar\omega/k_{B}T$ on a log-log scale.  The
data collapse onto a single curve.  The lines are fits as described
in the text.}\vspace{-6mm} \label{Scaling}
\end{figure}

The resulting values for $\chi''(\omega)$ follow a similar scaling
relation as the ac susceptibility, where the ratio
$\hbar\omega/k_{B}T$ replaces $\mu_{B}H/k_{B}T$. In
Fig.~\ref{Scaling} we show $\chi''(\omega)T^{0.66}$ on the $y$ axis
and the unitless ratio $\hbar\omega/k_{B}T$ on the $x$ axis. The
scaled data collapse fairly well onto a single curve over almost
four decades of $\hbar\omega/k_{B}T$.  Here we have used the same
exponent $\alpha = 0.66$ that was observed in the
scaling of the ac susceptibility.   However, the error bars on the
data allow for a wider range of exponents ($\alpha = 0.55$ to 0.75)
with reasonable scaling behavior. The collapse of the
$\chi''(\omega)$ data is again reminiscent of the behavior observed
in certain heavy-fermion metals, including the shape of the
functional form of the scaling function. Let us assume that
$\chi''(\omega)T^{\alpha}~\propto~F(\omega/T)$. The
heavy-fermion metal CeCu$_{5.9}$Au$_{0.1}$ displays a
scaling\cite{Schroder,SchroderNature} that could be fit to the
functional form $F(\omega/T) =  sin[\alpha \, tan^{-1}(\omega/T)]/[(\omega/T)^{2}  +  1]^{\alpha/2}$.  A fit
to this functional form is shown as a dashed blue line in
Fig.~\ref{Scaling}.  This simple form does not fit the
herbertsmithite data well for low values of $\omega/T$.  Other
heavy-fermion metals\cite{Park,Aronson1995}, display a scaling
relation that can be fit to the functional form $F(\omega/T)  =
(T/\omega)^{\alpha}tanh(\omega/\beta T)$; this functional form is
similar to that used to fit the dynamic susceptibility in
La$_{1.96}$Sr$_{0.04}$CuO$_4$\cite{Keimer}. This functional form
fits our data much better, shown (with fit parameter $\beta = 1.66$) as the dark red line in
Fig.~\ref{Scaling}. This function is somewhat unusual, in that for
low values of $\omega/T$ it is proportional to
$(\omega/T)^{1-\alpha}$ rather than the expected $\omega/T$\cite{SachdevYe}; of course such a dependence might
be recovered at still smaller values of $\omega$. For larger values
of $\omega/T$, this curve approaches a power law dependence with
$\chi''(\omega)  \propto  \omega^{-\alpha}$. This is consistent
with the low temperature ($T=35$ mK) behavior of the dynamic
susceptibility of herbertsmithite reported earlier\cite{Helton}.

Other works on kagome lattice systems have shown evidence for
similar behavior of the susceptibility.  The dynamic susceptibility
in the kagome bilayer compound SCGO has been shown to display power law behavior\cite{Broholm} and has been fit
to a form\cite{Mondelli} identical to that shown as the dark red line in Fig.~\ref{Scaling}
with $\alpha = 0.4$. Both SCGO and BSGZCO\cite{Bonnet}
demonstrate anomalous power law behavior in their bulk
susceptibilities.  Also, an early dynamical mean-field
theory study of a kagome RVB state\cite{Georges} predicted such
a scaling of the dynamic susceptibility.  A recent paper\cite{deVries} on herbertsmithite
found that $S(\omega)$ was roughly independent of both temperature and
energy transfer for values of $\omega$ greater than 2 meV.  This
simpler $\omega/T$ scaling is different from what we measure here in the
low energy susceptibility.

Similar scaling has been reported in other quantum antiferromagnets,
many of which are believed to be close to a quantum phase
transition\cite{SachdevYe}. The neutron scattering
results on the spin-glass La$_{1.96}$Sr$_{0.04}$CuO$_4$ show a
scaling that at small values of $\omega$ worked best with $\alpha=
0.41 \pm 0.05$\cite{SachdevYe},  while at higher energy transfers
the data followed a pure $\omega/T$ scaling\cite{Keimer} with
$\alpha$ = 0.  Further comparisons can be made to neutron results on
various heavy-fermion metals with doping levels that place them near
a transition to a spin-glass or antiferromagnetically ordered
state\cite{Aronson1995,Schroder,Wilson,Park}.  In
addition, a scaling of the ac susceptibility similar to that
reported here was seen in
CeCu$_{5.9}$Au$_{0.1}$\cite{SchroderNature} and
Ce(Ru$_{0.5}$Rh$_{0.5}$)$_{2}$Si$_{2}$\cite{Tabata2004}.  Recent
exact diagonalization work\cite{Cepas} has suggested that
the ground state of the spin-$\frac{1}{2}$ kagome lattice
antiferromagnet with a Dzyaloshinskii-Moriya interaction will be a
quantum disordered state for the Heisenberg Hamiltonian, but a
N\'{e}el ordered state when the component of the
Dzyaloshinskii-Moriya vector perpendicular to the kagome lattice
plane exceeds $\approx J/10$.  The presence of a nearby QCP is also
possible in models without a DM interaction\cite{Poilblanc}.
Futhermore, several of the theoretically proposed ground states for
the spin-$\frac{1}{2}$ kagome lattice
antiferromagnet\cite{Ran,Ryu} are critical or algebraic spin
liquid states.  These proposed quantum ground states
would possibly display excitations that are similar to
fluctuations near a QCP.  Thus, the observed low energy scaling
behavior in herbertsmithite might signify quantum critical
behavior\cite{SachdevNP} or a critical
spin liquid ground state.

In many doped heavy-fermion metals, the observed non-Fermi liquid
behavior is likely related to disorder. In
herbertsmithite, the low temperature susceptibility roughly
resembles a Curie tail, and it has been
suggested\cite{Misguich,deVries1} that this is attributable to
$S=1/2$ impurities (consisting of $\approx 5$\% of all magnetic ions)
with weak couplings to the rest of the system.  We find that the
scaling behavior seen in herbertsmithite does have features in
common with the disordered heavy-fermion metals, such as
$\chi''(\omega)$ proportional to $(\omega/T)^{1-\alpha}$ at low
values of $\omega/T$ rather than linear in $\omega/T$.  The
divergence of the low temperature susceptibility may also be
indicative of a random magnetic system, such as a Griffiths
phase\cite{CastroNeto1998,Griffiths} or random singlet phase\cite{BhattLee}.  A
collection of impurity spins subject to a broad distribution of
couplings, $P(J_{imp})$ may result in a power law susceptibility
at sufficiently low temperatures\cite{Ma}; however, the scaling presented here describes the entire measured susceptibility rather than the response of a small impurity fraction. In this scenario, the scaling of the ac
susceptibility, with $\chi^{\prime}T^{0.66}  \propto  F(H/T)$,
would be useful in determining the distribution of couplings
experienced by the impurity spins in herbertsmithite.  The
disordered heavy-fermion metal
Ce(Ru$_{0.5}$Rh$_{0.5}$)$_{2}$Si$_{2}$ displays a scaling of the ac
susceptibility\cite{Tabata2004} that is remarkably similar to that
in herbertsmithite, except with a considerably larger effective
moment [a much smaller field will suppress the low temperature
susceptibility of Ce(Ru$_{0.5}$Rh$_{0.5}$)$_{2}$Si$_{2}$].  That
scaling was attributed to a broad distribution of coupling
strengths that likewise diverges at low coupling, quite likely with
a power law distribution\cite{Liu}: $P(J_{imp})  \propto
J_{imp}^{-\alpha}$. In terms of our data, this would imply a
distribution of impurity couplings which extend to several meV,
which is surprisingly large for the assumed out-of-plane impurity
ions. A further prediction in this scenario is that the distribution
of local susceptibilities would diverge at low
temperatures\cite{CastroNeto1998} such that the width of the Knight
shift, $\delta K$, as measured by NMR would go as
$\frac{\delta K}{K} \propto  T^{-\lambda}$ with $\lambda \simeq 0.17$. The observed NMR signal
certainly broadens at low temperatures\cite{Imai}, and it
would be most interesting to see if it follows this specific power
law.  Thermal transport measurements would be important to help differentiate between scenarios where the scale-invariant spin excitations are localized near impurities or extended (as in the aforementioned criticality scenarios).

In conclusion, we have shown that the low energy dynamic
susceptibility of the spin-$\frac{1}{2}$ kagome lattice
antiferromagnet herbertsmithite displays an unusual scaling relation
such that $\chi T^{\alpha}$ with $\alpha \simeq 0.66$ depends only
on the thermal energy scale $k_{B}T$ over a wide range of
temperature, energy, and applied magnetic field. This behavior is
remarkably similar to the data seen in certain quantum
antiferromagnets and heavy-fermion metals as a signature of
proximity to a quantum critical point.  In addition to scenarios
based on impurities, the results may indicate that the
spin-$\frac{1}{2}$ kagome lattice antiferromagnet is near a QCP,
or that the ground state of herbertsmithite may behave like a
critical spin liquid.

We thank J.W. Lynn, C. Payen and T. Senthil for helpful discussions.  JSH
acknowledges support from the NRC/NIST Postdoctoral Associateship
Program.  The work at MIT was supported by the Department of Energy
(DOE) under Grant No.~DE-FG02-07ER46134.  This work used facilities
supported in part by the NSF under Agreement No.~DMR-0454672.\\ \\
$*$ email: younglee@mit.edu \\ \\
Current addresses: \\ a) Department of Physics, Mahidol University, Thailand \\ b) Department of Chemistry, Colorado State University \\ c) Department of Chemistry, University of Michigan

\bibliography{Scaling}
\end{document}